\documentstyle{article}\begin{document}
\title{Relativistic diffusion of quarks in random gluon fields}
\author{Z. Haba \\
Institute of Theoretical Physics, University of Wroclaw,\\ 50-204
Wroclaw, Plac Maxa Borna 9, Poland\\email:zhab@ift.uni.wroc.pl}
\date{}\maketitle
\begin{abstract} We consider Wong equations for a particle with a continuous mass spectrum in a random Yang-Mills field
 approximating the quantum field at finite temperature.
We show that   particle time  evolution can be approximated  by a
relativistic diffusion. Kubo's  generator of the relativistic
diffusion is defined as an expectation value of the square of the
Liouville operator.\end{abstract}

 \section{Introduction}The diffusion approximation is well-known
 in  non-relativistic physics \cite{pitaj}. It has been applied in a description of
 the quark-gluon plasma and in heavy
ion collisions
\cite{hwa}\cite{svetitsky}\cite{rafelski}\cite{rupp}. If QCD is
considered as the fundamental
 theory of  strong interactions then the diffusion should result from the QCD transport
 theory \cite{heinzprl}\cite{heinz}\cite{elze}. The form of
the diffusion operator can in principle be determined on the basis
of the Boltzmann kinetic equation after a calculation of the
scattering amplitude \cite{svetitsky}\cite{groot}. A
semi-classical approach to
 the transport theory based on the Wong equation \cite{wong}
( derived from quantum dynamics in
\cite{brown}\cite{taylor}\cite{lit1})
 leads to an analog of the Vlasov equation \cite{heinzprl}\cite{heinz}
 \cite{lit1}\cite{lit2}. Quantized gauge fields
in the Vlasov equation have been discussed in
\cite{gul1}\cite{gul2}\cite{tean}. These authors have shown that
quark's time evolution in a quantum gauge field leads to a
diffusion of momentum and color. However, the generator of the
diffusion has not been defined in an unambiguous way ( the
coefficients of the diffusion equation are undetermined and the
relativistic invariance remains obscure). It is well-known
\cite{to}\cite{plasma} that the Vlasov equation
 describing a tracer particle  moving in
 plasma in a random electromagnetic field produced by a chaotic
 motion of other particles
  can be approximated by a diffusion equation.
The generator of the diffusion has been calculated in
\cite{to}\cite{plasma}. A detailed discussion of the diffusion
approximation to a random non-relativistic dynamics can be found
in \cite{kubo1}\cite{kubo2} and \cite{kesten}.

There is no consensus on the relativistic form of the diffusion
(see the reviews in \cite{hang}\cite{deb}). A relativistic
diffusion in the configuration space does not exist
\cite{lopuch}\cite{hakim}. If we assume that the diffusing
particle has a fixed mass then an analog of the Kramers diffusion
(in momentum space) is uniquely defined by its generator being the
Laplace-Beltrami operator on the mass-shell as shown  by Schay
\cite{schay} and Dudley \cite{dudley} (for  further developments
in this approach see \cite{habapre}\cite{calo}). There are some
other suggestions for a definition of the relativistic diffusion
without the assumption that the diffusion takes place on the
mass-shell \cite{hang}\cite{deb}\cite{deb2}.

In this paper we start from the basic principles of quantum gauge
theory at finite temperature. We assume the gauge and Lorentz
covariance in the most general form. It is  known that the
thermodynamic equilibrium is defined in the rest frame. Then, if
the finite temperature theory is Lorentz covariant then it can be
transformed to an arbitrary Lorentz frame described by a unit
time-like four-vector $w^{\mu}$. We consider the two-point
function of gauge covariant field strength \cite{bial}\cite{man}
$\hat{F}_{\mu\nu}(x)$. We show that the generator of the diffusion
of momentum  depends only on the form of the two-point function of
$\hat{F}_{\mu\nu} $ at coinciding points. The diffusion tensor can
be related to the energy-momentum tensor of gluons in the thermal
state. We do not assume that the particles (quarks) are on the
mass-shell as we did in the case of (zero-temperature) QED in
\cite{habajpa}. Only free stable particles satisfy the mass-shell
condition. Relaxing this condition allows an explicit use of  the
Lorentz covariance ( four independent components of the
momentum;the transport theory for particles with a continuous mass
spectrum has been discussed also in ref.\cite{mrow}). In this
paper we extend the covariant calculation of the diffusion in
momentum space derived for QED at finite temperature in
\cite{habajpa2} to QCD. For this purpose we apply the Wong
approximation for quark dynamics in QCD. The plan of the paper is
the following. In sec.2 we formulate the Wong equation in the
random field. In sec.3 we calculate the expectation value of the
random evolution (the square of the Liouville operator) which
determines the diffusion generator. We discuss the resulting
diffusion equation and its applications.

\section{Kinetic equation of Wong dynamics in a  Yang-Mills field}
We consider a classical approximation to quantum gauge theory when
quark evolution is described by classical trajectories  and gluons
are approximated by classical Yang-Mills fields. Such an
approximation can be justified in the regime of low external
momenta and high temperature. In this regime the approximation
leads to the Wong equation \cite{wong} which can be derived from
QCD quantum dynamics \cite{brown}\cite{taylor}\cite{lit2}.

 In Wong dynamics we have an extended phase space: a product   ${\cal P}\times
R^{n}$ of the usual phase space ${\cal P}$ of position and momenta
$(x(\tau),p(\tau))$  and the color  described by $n$ coordinates
$Q^{a}$. The evolution parameter $\tau$   is chosen as proportional
to the proper time.
 The dynamics of a particle
in a Yang-Mills field $A_{\mu}^{a}$ is described by the equations
\cite{wong}
 \begin{equation}
\sqrt{p^{2}} \frac{dx^{\mu}}{d\tau}=p^{\mu},
 \end{equation}
 \begin{equation}
\sqrt{p^{2}}\frac{dp_{\mu}}{d\tau}=Q_{a}F^{a}_{\mu\nu}p^{\nu},
 \end{equation}

\begin{equation}
\frac{dQ_{a}}{d\tau}=-f^{abc}p^{\mu}A^{b}_{\mu}Q_{c},
\end{equation}
 where $\mu=0,1,2,3$ and $f^{abc}$ are the structure constants of the color group ${\cal G}$.
It follows from eq.(3) that \begin{equation}
\frac{d}{d\tau}Q^{2}\equiv\frac{d}{d\tau}(Q_{a}Q_{a})=0
\end{equation}and from eq.(2)
\begin{equation}
\frac{d}{d\tau}p^{2}\equiv\frac{d}{d\tau}(\eta^{\mu\nu}p_{\mu}p_{\nu})
=0,
\end{equation} where $\eta_{\mu\nu}=(1,-1,-1,-1)$. From eqs.(1)
and (5) it follows that
\begin{equation} d\tau^{2}=dx^{\mu}dx_{\mu}p^{-2}.
\end{equation}Hence, $d\tau $ is $c\sqrt{ p^{-2}}$ times the proper time $dt\sqrt{1-c^{-2}(\frac{d{\bf x}}{dt})^{2}}$ where $x^{0}=ct$ ( we
set $c=1$ from now on for convenience).

It will be useful to work with a matrix notation. Let $T_{a}$ be a
representation of the algebra of the group ${\cal G}$
\begin{displaymath}
[T_{a},T_{b}]=f^{abc}T_{c}
\end{displaymath}
normalized by $Tr(T_{a}T_{b})=-\delta_{ab}$. We define
$Q=Q_{a}T_{a}$ and $A_{\mu}=T_{a}A^{a}_{\mu}$. Let
$Ap=p^{\mu}A_{\mu}$ then eq.(3) can be expressed as
\begin{displaymath}
\frac{dQ}{d\tau}=-[Ap,Q]
\end{displaymath}
It has a solution
\begin{equation}
Q_{\tau}=U_{\tau}Q_{0}U_{\tau}^{-1}
\end{equation}
where $U_{\tau}\in {\cal G}$ solves the equation
\begin{displaymath}
dU_{\tau}=-pA U_{\tau}\end{displaymath} Eq.(7) is realizing the
sphere (4) as an orbit of the adjoint representation of ${\cal
G}$.
 We shall denote  the
invariant measure on the sphere $Q^{2}\equiv TrQ^{2}$ by $dQ$ (we
use capital Latin letters to denote the points on the sphere).

The kinetic theory of classical  particles  begins with a
distribution ${\cal K}_{\tau}$ of trajectories
$(x(\tau,y,q,Q),p(\tau,y,q,Q),R(\tau,y,q,Q))$
 starting from $(y,q,Q)$ in the  phase space
\begin{equation}\begin{array}{l}
{\cal
K}_{\tau}(x,p,S;y,q,Q)\cr=\delta(x-x(\tau,y,q,Q))\delta(p-p(\tau,y,q,Q))
\delta(S-R(\tau,y,q,Q)).\end{array}
\end{equation} We can consider more general
distributions of trajectories $\Phi_{\tau}$ by spreading the initial
points in the phase space with a certain probability distribution
$\Phi$
\begin{equation}
\Phi_{\tau}(x,p,R)=\int {\cal
K}_{\tau}(x,p,R;y,q,Q)\Phi(y,q,Q)dydqdQ,
\end{equation} (
by $dy$ or $dq$ we denote an integral over $R^{4}$). ${\cal
K}_{\tau}$ as well as $\Phi_{\tau}$ satisfy the differential
equation
\begin{equation}\begin{array}{l}\frac{d}{d\tau}\Phi_{\tau}\equiv {\cal L}^{+}\Phi_{\tau}=
-\frac{1}{\sqrt{p^{2}}}(p^{\mu}\frac{\partial\Phi_{\tau}}{\partial
x^{\mu}}+ F^{\mu\nu}p_{\nu}\frac{\partial\Phi_{\tau}}{\partial
p^{\mu}}+f^{abc}
R_{a}p^{\mu}A_{\mu}^{b}\frac{\partial\Phi_{\tau}}{\partial
R_{c}}).
\end{array}\end{equation} ${\cal K}_{\tau}$ solves eq.(10) with  the  initial condition
$\delta(x-y)\delta(p-q)\delta(S-Q)$ and  $\Phi_{\tau}(x,p,R)$ with
the initial condition $\Phi(x,p,Q)$ .  In eq.(10) we assumed that
$p_{0}$ is an independent variable. Keeping the four-momentum off
the mass shell allows to preserve explicit Lorentz invariance. The
off-shell formulation appears in the Wigner function approach to
quantum theories  \cite{groot}\cite{elze}\cite{mrow} (the
mass-shell projection is performed as an additional step).

The density of trajectories $\Omega$ in the laboratory time
$x^{0}$ can be expressed by $\Phi_{\tau}$ \cite{groot}
\begin{equation}
\Omega(x,p,R)=\int d\tau \Phi_{\tau}(x,p,R).
\end{equation}
It satisfies the transport equation \begin{equation} {\cal
L}^{+}\Omega=0.
\end{equation}In the Liouville approach to the classical statistical
mechanics we consider functions $W$ on the phase space as
observables. We may define the expectation value $\Phi_{\tau}(W)$ of
$W$ in the state $\Phi_{\tau}$ as
\begin{equation}\begin{array}{l}\Phi_{\tau}(W)=
(\Phi_{\tau},W)=\int
dxdpdQ\sigma(p)\Phi_{\tau}(x,p,Q)W(x,p,Q)\equiv(\Phi,W_{\tau})\cr=
\int
dydqdQ\sigma(q)\Phi(y,q,Q)W(x(\tau,y,q,Q),p(\tau,y,q,Q),R(\tau,y,,p,Q)).\end{array}
\end{equation} Here, $\sigma(p)$ is a mass distribution depending on $p^{2}$.
In order to fix the mass we let $\sigma(p)\rightarrow
\delta(p^{2}-m^{2})$. The semigroup $(T_{\tau}W)(y,q,P)\equiv
W_{\tau}(y,q,P)=W(x(\tau,y,q,P),p(\tau,y,q,P),R(\tau,y,q,P))$ is
unitary in the scalar product (13). We have from eq.(13)
\begin{equation}\begin{array}{l}\frac{d}{d\tau}W_{\tau}(y,q,P)=
\frac{d}{d\tau}(T_{\tau}W)(y,q,P)=({\cal L}T_{\tau}W)(y,q,P) \cr=
\frac{1}{\sqrt{q^{2}}}(q^{\mu}\frac{\partial W_{\tau}}{\partial
y^{\mu}}+F^{\mu\nu}q_{\nu}\frac{\partial W_{\tau}}{\partial
q^{\mu}}+f^{abc} P_{a}p^{\mu}A_{\mu}^{b}\frac{\partial
W_{\tau}}{\partial P_{c}}),
\end{array}\end{equation}where the generator ${\cal L}$ of the
semigroup $T_{\tau}=\exp(\tau{\cal L})$ is the adjoint of ${\cal
L}^{+}$ of eq.(10) in the Hilbert space $L^{2}(\sigma dydqdP)$ .
The requirement of the $\tau$-independence of the probability
distribution ($\frac{d}{d\tau}\Phi_{\tau}=0$) gives the kinetic
equation in the laboratory time $t$ . This is the same equation as
the one which can be derived by an elimination of $\tau$ in favor
of $t$ in the evolution equations (1)-(2).

  Let us consider the
equation (14) for functions $W$ of particle trajectories which is
of the form
\begin{equation}
\partial_{\tau}W=(X+Z+Y)W\equiv(K+Y)W,
\end{equation}
where $K=X+Z$. In eq.(15) we have according to eq.(14)
\begin{equation} X=p^{\mu}\partial^{x}_{\mu},\end{equation}\begin{equation}\begin{array}{l}
Y=-Tr (QF_{\mu\nu}(x)p^{\nu}\frac{\partial}{\partial
p^{\mu}},\end{array}\end{equation}
\begin{equation}Z=-
Tr\Big([ Q,p^{\mu}A_{\mu}(x)]\frac{\partial}{\partial Q}\Big).
\end{equation}Let
\begin{equation}
Y(s)=\exp(-sK)Y\exp(sK). \end{equation} Then, the solution of
eq.(15) can be expressed as (an analog of the interaction picture)
\begin{equation}
W_{t}=\exp(tK)W^{I}_{t},
\end{equation}
where $W_{t}^{I}$ is a solution of the equation
\begin{equation}
\partial_{s}W^{I}_{s}=Y(s)W^{I}_{s}.
\end{equation}
In order to calculate $Y(s)$ we introduce  the path-ordered
exponential along a curve $\gamma$ starting from $x$
\begin{equation}
U_{\gamma y}=P\Big(\exp(-\int A_{\mu}d\gamma^{\mu})\Big).
\end{equation}
with $y$ as the endpoint of the curve. It is a solution of the
equation
\begin{equation}
\frac{\partial}{\partial y^{\mu}}U_{\gamma y}=-A_{\mu}(y)U_{\gamma
y}.
\end{equation}
Under the gauge transformation
\begin{displaymath}
A_{\mu}\rightarrow gA_{\mu}g^{-1}-g\partial_{\mu}g^{-1}
\end{displaymath}we have $F_{\mu\nu}(y)\rightarrow
g(y)F_{\mu\nu}(y)g(y)^{-1}$ and
\begin{displaymath}U_{\gamma y}\rightarrow g(y)U_{\gamma
y}g^{-1}(x) \end{displaymath}  In terms of the path ordered
exponential the solution of eq.(3) is
\begin{equation} Q(\tau)=U_{p}(\tau)Q(0)U_{p}^{-1}(\tau),
\end{equation} where $U_{p}(\tau)=U_{\gamma y}$ is  calculated  along the
 straight line in the direction $p$,i.e., $\gamma(s)=x-ps$
 with $y=x-p\tau$ .

We define the corresponding gauge covariant field strength
transported along the straight line \cite{bial}\cite{man}
\begin{equation}
\hat{F}_{\mu\nu}(p,x-ps)=U_{p}(s)^{-1}F_{\mu\nu}(x-ps)U_{p}(s).
\end{equation}Under a gauge transformation\begin{equation}
\hat{F}_{\mu\nu}(p,x-ps)\rightarrow
g(x)\hat{F}_{\mu\nu}(p,x-ps)g(x)^{-1}
\end{equation}
(here $x$ is a fixed point, so the gauge rotation does not change
with $s$).
 From
eqs.(17) and (19) (using the Hausdorf-Baker formula) we obtain
\begin{equation}\begin{array}{l} Y(s)=-Tr
\Big(Q(s)F_{\mu\nu}(x-sp)\Big)p^{\nu}(\frac{\partial}{\partial
p_{\mu}}+s\frac{\partial}{\partial x_{\mu}})+sf^{abc} Q_{a}A^{b}_
{\rho}Q_{d}F_{d}^{\sigma\rho}p_{\sigma}\frac{\partial}{\partial
Q_{c}}+o(s^{2})\cr=-Tr
\Big(Q(0)\hat{F}_{\mu\nu}(p,x-sp)\Big)p^{\nu}(\frac{\partial}{\partial
p_{\mu}}+s\frac{\partial}{\partial x_{\mu}})+sf^{abc} Q_{a}A^{b}_
{\rho}Q_{d}F_{d}^{\sigma\rho}p_{\sigma}\frac{\partial}{\partial
Q_{c}}+o(s^{2}).\end{array}\end{equation} The terms linear in $s$
as well as the $o(s^{2})$ terms (higher orders in $s$) will be
irrelevant for the calculation of the diffusion performed in the
next section.

The assumption that the field strength $F$ is constructed from
$A_{\mu}$is expressed as the Bianchi identities
\begin{displaymath}
 \epsilon^{\mu\nu\sigma\rho}D_{\mu}F_{\sigma\rho}=0,\end{displaymath}
where
$D_{\mu}F_{\sigma\rho}=\partial_{\mu}F_{\sigma\rho}+[A_{\mu},F_{\sigma\rho}]$.
We consider random gluon fields as an approximation to quantum
fields. In general, correlation functions of gauge fields being
gauge dependent are not well-defined.  We are going to define a
two-point correlation function of gauge covariant variables
\begin{equation}
\hat{F}_{\mu\nu}(\gamma,y)=U^{-1}_{\gamma y}F_{\mu\nu}(y)U_{\gamma
y}.
\end{equation}
$\hat{F}$ satisfies the Bianchi identities with an ordinary
derivative (instead of the covariant one;this has been shown first
in \cite{bial}, see also \cite{man})\begin{equation}
 \epsilon^{\mu\nu\sigma\rho}\partial_{\mu}\hat{F}_{\sigma\rho}=0\end{equation}

We treat the Wong equations (1)-(3) in a random Yang-Mills field
as an approximate description of the dynamics of quarks in quantum
gauge  theory at high temperature. Then, the expectation value of
an observable ${\cal O}$ at the inverse temperature $\beta$ is
defined as
\begin{equation}
\langle {\cal O}\rangle_{\beta}=Tr\Big(\exp(-\beta
w_{\mu}P^{\mu}){\cal O}\Big)
\end{equation}
where $P^{\mu}$ is the generator of space-time translations and
the  four-vector $w^{\mu}$ ($w^{\mu}w_{\mu}=1$) describes the
reference frame \cite{mat}.

In QED (and perturbative QCD) the expectation value (30) leads to
the distribution $ n(p;w)$ of photons (gluons)
 according to the covariant Bose-Einstein law
\begin{equation}
n(p;w)=(\exp(\beta w^{\nu}p_{\nu})-1)^{-1}
\end{equation}
\section{Diffusive motion in a random gauge field}

 The two-point correlation functions of gauge
covariant variables $\hat{F}(p,x-ps) $ are covariant under
transformations of the Poincare group (acting upon $x$,$p$ and
$w$) and invariant under transformations of the (color) ${\cal G}$
symmetry. We assume that the non-commutativity of quantum gauge
fields can be neglected at high temperature in the calculation of
the correlation functions (this can be shown in perturbation
theory in the state (30)). Then, the symmetry properties with
respect to an exchange of indices together with the Bianchi
identities (29) lead to the representation ( the Bianchi
identities in the form (29) allow to apply the same methods for
the two-point function as in the Abelian case of \cite{habajpa2})

\begin{equation}\begin{array}{l}
\langle
\hat{F}^{a}_{\mu\nu}(p,x-ps))\hat{F}^{b}_{\sigma\rho}(p,x-ps^{\prime})\rangle
\equiv\delta^{ab}G^{p}_{\mu\nu
;\sigma\rho}(p(s^{\prime}-s);s,s^{\prime},w),\end{array}\end{equation}
 where \begin{equation}\begin{array}{l}
 G^{p}_{\mu\nu
;\sigma\rho}(y;s,s^{\prime},w)\cr= \int
dk\tilde{G}(s,s^{\prime},k,p,w)(\eta_{\mu\sigma}k_{\nu}k_{\rho}-\eta_{\mu\rho}k_{\nu}k_{\rho}
+\eta_{\nu\rho}k_{\sigma}k_{\mu}-\eta_{\nu\sigma}k_{\mu}k_{\rho})\exp(iky).
\end{array}\end{equation}  $\tilde{G}$ is a Lorentz invariant
function of $k$, $p$ and $w$. Let us note that
\begin{equation} G^{p}_{\mu\nu
;\sigma\rho}(0;0,0,w)=\eta_{\mu\sigma}T_{\nu\rho}-\eta_{\mu\rho}T_{\nu\sigma}
+\eta_{\nu\rho}T_{\sigma\mu}-\eta_{\nu\sigma}T_{\mu\rho}
\end{equation}where
\begin{equation}
T_{\mu\nu}=\int dk\tilde{G}(0,0,k,0,w)k_{\mu}k_{\nu}=\rho
w_{\mu}w_{\nu}-\pi_{E}(\eta_{\mu\nu}-w_{\mu}w_{\nu})
\end{equation}Eq.(35) gives a decomposition of the tensor
$T_{\mu\nu}$ in terms of $\eta_{\mu\nu}$ and $w_{\mu}w_{\nu}$. If
$\tilde {G}$ is non-negative then $\rho$ and $\pi_{E}$ must be
non-negative. $\tilde{G}$ at $s=s^{\prime}=0$ does not depend on
$p$ because $p$ can enter $F$ only through the path starting in
$0$ and ending in $sp$. $T_{\mu\nu}$ is related to the expectation
value (32) of the Yang-Mills energy-momentum tensor. In order to
see this let us calculate
\begin{equation}\begin{array}{l}
-Tr\langle
F_{\mu\rho}F_{\nu\sigma}\eta^{\rho\sigma}-\frac{1}{4}\eta_{\mu\nu}
F_{\sigma\rho}F^{\sigma\rho}\rangle=2T_{\mu\nu}-\frac{1}{2}\eta_{\mu\nu}T_{\rho}^{\rho}
\end{array}\end{equation}
We are going to approximate the evolution (15) by a diffusion in
momentum space. As follows from eqs.(1)-(3) such an evolution is
covariant but not invariant with respect to the gauge
transformations. We show that the diffusion in the momentum space
can be defined in a gauge invariant way. The color also evolves in
time. However, this evolution is gauge-dependent. We give
arguments at the end of this section that if the color evolution
is approximated by diffusion then there is a gauge such that the
color diffusion constant is equal to zero.

We apply the  approach of Kubo \cite{kubo1}-\cite{kubo2} which
approximates the random Liouville operator on the rhs of eq.(10)
by the expectation values of $Y$ and its square
\begin{equation}
\langle W_{\tau}^{I}\rangle \simeq \exp\Big(\int_{0}^{\tau}ds\langle
Y(s)\rangle+\frac{1}{2}\langle(\int_{0}^{\tau}ds
\tilde{Y}(s))^{2}\rangle\Big)W
\end{equation}
 where
\begin{equation}
\tilde{Y}=Y-\langle Y \rangle
\end{equation}
The second order term resulting from the square of $Y$ reads ( we
omit the terms with the derivatives over the charges $Q$
restricting ourselves to the diffusion in momentum; we assume that
the initial state $W$ in eq.(15) does not depend on $Q$)
\begin{equation}\begin{array}{l}\lim_{\tau\rightarrow
0}\frac{1}{\tau^{2}}\langle(\int_{0}^{\tau}Y(s)ds)^{2}\rangle =
\lim_{\tau\rightarrow
0}\frac{1}{2\tau^{2}}\int_{0}^{\tau}ds\int_{0}^{\tau}ds^{\prime}\langle
Y(s)Y(s^{\prime})+Y(s^{\prime})Y(s)\rangle \cr=p^{-2}Q^{2}
(\eta_{\mu\sigma}T_{\nu\rho}-\eta_{\mu\rho}T_{\nu\rho}
+\eta_{\nu\rho}T_{\sigma\mu}-\eta_{\nu\sigma}T_{\mu\rho})p^{\nu}\partial^{\mu}
p^{\rho}\partial^{\sigma}\equiv -{\cal A}_{w}.
\end{array}\end{equation}
Elementary calculations using eq.(34) give
\begin{equation}\begin{array}{l}
{\cal A}_{w}=-2\pi_{E}Q^{2}\partial_{\mu}P^{\mu\nu}\partial_{\nu}
+Q^{2}(\rho+\pi_{E})p^{-2}((wp)^{2}\partial_{\mu}\partial^{\mu}-2wpw^{\mu}p^{\rho}\partial_{\mu}\partial_{\rho}
\cr+p^{2}w^{\mu}w^{\rho}\partial_{\mu}\partial_{\rho}-w^{2}p^{\nu}\partial_{\nu}
-2wpw^{\nu}\partial_{\nu})\equiv
\partial_{\mu}\alpha^{\mu\nu}\partial_{\nu},
\end{array}\end{equation}
 where
 \begin{displaymath}
 P^{\mu\nu}=\eta^{\mu\nu}-p^{-2}p^{\mu}p^{\nu}\end{displaymath}
  and $Q^{2}\equiv Tr Q^{2}=Q_{a}(\tau)Q_{a}(\tau)=Q_{a}(0)Q_{a}(0)$ is the square of
the charge (which is a constant according to eq.(7)). The term
$s\partial_{\mu}^{x}$ of eq.(27) as well as the remaining terms
linear in $s$ and $o(s^{2})$  do not contribute to ${\cal
A}_{w}$.This is so because it is of higher (third) order in
$\tau$. The limit in eq.(40) depends only on $G(0)$ corresponding
to $s=s^{\prime}=0$ because taking into account $s$ in $G$ would
give higher orders in $\tau$. As a consequence the $T_{\mu\nu}$
coefficients  depend neither on $x$ nor on $p$ because the
dependence on $p$ is always multiplied by $s$. According to Kubo
\cite{kubo1}-\cite{kubo2} the $\tau^{2}$ behavior at short time
(small in comparison to the correlation time $\tau_{c}$) of the
random evolution goes over to the linear behavior at long time
with the generator ${\cal A}_{w}$ multiplied by the correlation
time $\tau_{c}$ of the gluon modes.
 In order to justify a replacement of the Liouville evolution (15)
  by the diffusion (40) at an arbitrarily long time
  we have to apply the Markov approximation. In such a case
  the dynamics is viewed as a composition of independent short time
  evolutions starting anew after the correlation time $\tau_{c}$
  (see the discussion in Kubo \cite{kubo1}-\cite{kubo2} ). The rigorous proof \cite{kesten}
requires a proper scaling of fields and the time in order to
determine precisely the conditions required for the limiting
diffusive behavior.

We define the diffusion generator by the lowest order (
$\tau^{2}$) term in the expansion (37) in proper time. We
incorporate Kubo's argument that this operator multiplied by the
correlation time generates a diffusion properly approximating the
random dynamics. In this way we are led to the following diffusion
equation for particles with a continuous mass spectrum ( by a
certain abuse of notation we denote by $W$ the expression which
has the meaning of $\langle W\rangle $ in random dynamics of
sec.2)
\begin{equation}\begin{array}{l}
\frac{1}{\tau_{c}}\partial_{\tau}W=\frac{1}{\tau_{c}}p^{\mu}\partial^{x}_{\mu}W+\partial_{\mu}\alpha^{\mu\nu}\partial_{\nu}W.
\end{array}\end{equation} The
diffusion in the proper time originates from the relativistic
dynamics in the proper time. Then, the proper time has a
well-defined meaning. This is the time measured in the frame
moving with the particle. However, if the particle motion is
random then this frame is random as well. Such a random frame is
loosing a physical meaning. For this reason  we should express the
kinetic transport in terms of the coordinate (laboratory) time.
The probability distribution evolves according to the adjoint
equation (10). The independence of the proper time parametrization
($\frac{d}{d\tau}\Phi_{\tau}=0$) leading to the transport equation
(12) is equivalent to the replacement of the proper time by the
physical time $t$ in the kinetic equation (12). We extend this
requirement to the diffusion (41). In this way we obtain the
transport equation

\begin{equation}\begin{array}{l}
\frac{1}{\tau_{c}}p^{\mu}\partial^{x}_{\mu}\Omega=\partial_{\mu}\alpha^{\mu\nu}\partial_{\nu}
\cr=Q^{2}\partial_{\mu}\Big(2\pi_{E} P^{\mu\rho}\partial_{\rho}
-(\rho+\pi_{E})p^{-2}((pw)^{2}\partial^{\mu}-(wp)(w^{\mu}p^{\rho}
+w^{\rho}p^{\mu})\partial_{\rho}+p^{2}w^{\mu}w^{\rho}\partial_{\rho}\Big)\Omega.
\end{array}\end{equation}
 We have shown \cite{habajpa2} that the positivity condition
\begin{equation}a_{\mu}a_{\nu}\alpha^{\mu\nu}\geq 0 \end{equation}
 is satisfied under
the  condition
\begin{equation}\rho\geq \pi_{E}\geq 0.\end{equation}

We can  decompose the diffusion (42) into diffusions on the mass
shell using the formula
\begin{equation}
\sigma(p)=\int dm^{2}\sigma(m^{2})\delta(p^{2}-m^{2})
\end{equation} Then, at zero temperature (realized by setting  $\pi_{E}+\rho=0$ in eq.(42)) we have
\begin{equation}
-\int dp\sigma(p)\Phi \partial_{\mu}p^{2}P^{\mu\nu}\partial_{\nu}W
=\int dpdm^{2}\sigma(m^{2})\delta
(p^{2}-m^{2})\Phi\triangle_{H}^{m}W
\end{equation}
where \begin{equation}\begin{array}{l}
\triangle_{H}^{m}=(\delta^{jl}m^{2}+p^{j}p^{l})\frac{\partial}{\partial
p^{l}}\frac{\partial}{\partial
p^{j}}+3p^{l}\frac{\partial}{\partial p^{l}}\end{array}
\end{equation}$j,l=1,2,3.$ and  $\triangle_{H}^{m}$ is the generator of the relativistic diffusion
defined by Schay\cite{schay} and Dudley \cite{dudley}.The
relativistic invariance is fixing the diffusion coefficients in
front of the $\delta^{jl}$ and $p^{j}p^{l}$ terms which remained
undetermined in \cite{rafelski} and \cite{gul1}\cite{gul2}.

We did not discuss the charge evolution assuming that the initial
state $W$ does not depend on the charge $Q$. We could apply the
diffusion approximation to the random charge evolution determined
by the fluctuations of $A_{\mu}$ in eq.(3). If we choose the
coordinate gauge \cite{durand} $y^{\mu}A_{\mu}(x+y)=0$ then
\begin{equation}
A_{\mu}^{a}(x-ps)=\int_{0}^{1}d\lambda F^{a}_{\mu\nu}(x-\lambda
sp)sp^{\nu}.
\end{equation}
The diffusion constant evaluated  from
$\langle\frac{dQ^{a}}{d\tau}\frac{dQ^{b}}{d\tau^{\prime}}\rangle$
(using eq.(3)) along the straight line $y=sp$ is zero. This
calculation indicates that the color diffusion calculated in
\cite{gul1}-\cite{gul2} may be a spurious phenomenon.

The diffusion (42) determines the phase space distribution W. In
terms of $W$ we can define a velocity of the fluid of diffusing
particles and its energy-momentum tensor. Such an approach leads
to a hydrodynamic description \cite{rom} resulting directly from
QCD (in the Wong approximation).
\section{Summary}
We have  formulated the problem of  a transport theory of
quark-gluon plasma  starting from quantum gauge theory at finite
temperature in semiclassical approximation which is applicable at
high temperature and low momenta. In this approximation the quarks
are treated as classical color particles and gauge fields as
classical random fields. In this paper we have shown that the
diffusion in momentum results from a  model of a particle moving
in a gauge field in the same way as a non-relativistic diffusion
results from an evolution in a random electric and magnetic fields
\cite{to} \cite{plasma}.  The Wong equation has the gauge coupling
constant as the only parameter. Then, we derive the diffusion
equation which depends on two parameters $\rho$ and $\pi_{E}$
which could be calculated from the correlation functions in the
Euclidean framework on the lattice (the diffusion constant has
been expressed by correlation functions of Wilson loops in finite
temperature QCD in an earlier paper \cite{tean}) . The diffusion
equation  could be directly compared with experimental data ( see
ref. \cite{rupp}) or we could use it for some other
approximations. We have in particular in mind the hydrodynamic
equations which could be derived from the diffusion equation. The
parameters entering the fluid equations could be expressed by the
ones of the diffusion equation.


\begin{thebibliography}{99}\bibitem{pitaj} E.M. Lifshits and L.P. Pitaevskij,
Physical Kinetics,Pergamon Press,1981
\bibitem{hwa}R.C.
Hwa,Phys.Rev.{\bf D32},637(1985)
\bibitem{svetitsky} B.Svetitsky, Phys.Rev.{\bf D37},2484(1988)
\bibitem{rafelski}D.B. Walton
and J. Rafelski,Phys.Rev.Lett.{\bf 84},31(2000)
\bibitem{rupp}R.Rapp and H. van Hees,in Quark-Gluon Plasma 4, ed.by R.C.Hwa
and Xin-Nian Wang,World Scientific,2010; arXiv:0903.10961
\bibitem{heinzprl}U. Heinz, Phys.Rev.Lett.{\bf 51},351(1983)

\bibitem{heinz}U. Heinz, Ann.Phys.{\bf 161},48(1985)

\bibitem{elze} H.-Th. Elze and U. Heinz,
Phys.Rep.{\bf 183},81(1989)
\bibitem{groot}S.R. de Groot, W.A. van Leeuwen and Ch.G. van Weert,  Relativistic
Kinetic Theory, North Holland,1980
\bibitem{wong}S.K. Wong, Nuovo Cimento,{\bf A65},689(1975)
\bibitem{brown}L.S. Brown and W.I. Weisberger, Nucl.Phys.{\bf B157},285(1979)
\bibitem{taylor}J. Frenkel and J.C.
Taylor, Nucl.Phys.{\bf B334},199(1990)\bibitem{lit1}P.F.
Kelly,Q.Liu, C. Lucchesi and C. Manuel,
 Phys.Rev.{\bf D50},4209(1994)
\bibitem{lit2}D.F. Litim and C. Manuel,Phys.Rep.{\bf
364},451(2002)\bibitem{gul1}A.Selikhov and M. Gyulassy,
Phys.Lett.{\bf B316},373(1993)
\bibitem{gul2}A.Selikhov and M. Gyulassy, Phys.Rev.{\bf
C49},1726(1994)\bibitem{tean}J. Casalderrey-Solana and D.Teaney,
Phys.Rev.{\bf D74},085012(2006)\bibitem{to}W.B. Thompson and J.
Hubbard, Rev.Mod.Phys.{\bf 32},714(1960)
\bibitem{plasma}S. Ichimaru, Basic Principles of Plasma Physics,
 Benjamin-Cummings, London, 1973







\bibitem{kubo1}R.Kubo, J.Math.Phys.{\bf 4},174(1962)
\bibitem{kubo2}R. Kubo, M.Toda and N.Hashitsume, Statistical
Physics II.Nonequilibrium Statistical Mechanics, Springer,
Berlin,1985\bibitem{kesten}H. Kesten and G.C. Papanicolaou,
Commun.Math.Phys.{\bf 78},19(1980)
\bibitem{hang}J. Dunkel and P. H\"anggi,Phys.Rep.{\bf 471},1(2009)

\bibitem{deb}C. Chevalier and F. Debbasch,
 AIP Conf.Proc.{\bf
913},42(2007)
\bibitem{lopuch}J. Lopuszanski, Acta Phys.Polon.{\bf 12}, 87(1953)
\bibitem{hakim}R. Hakim, J.Math.Phys.{\bf 9},1805(1968)
\bibitem{schay} G.Schay,PhD thesis,Princeton University,1961
\bibitem{dudley} R.Dudley, Arkiv for Matematik,{\bf 6},241(1965)
\bibitem{habapre}Z. Haba, Phys.Rev.{\bf E79},021128(2009)
\bibitem{calo}J.A. Alcantara and S. Calogero, Kinetic and Related
Models, {\bf 4},401(2011)


\bibitem{deb2}C. Chevalier and F. Debbasch, J.Math.Phys.{\bf 49},043303(2008)


\bibitem{bial}I.
Bialynicki-Birula, Bull.Acad.Polon.Sci.{\bf 11},135(1963)



\bibitem{man}S. Mandelstam, Phys.Rev.{\bf 175},1580(1968)
\bibitem{habajpa}Z. Haba, Journ.Phys.{\bf A44},335202(2011)
\bibitem{mrow}S.Mrowczynski, Ann.Phys.(N.Y.){\bf 169},48(1986)\
\bibitem{habajpa2}Z. Haba, Journ.Phys.{\bf A46},155001(2013)
\bibitem{mat}T. Matsui, B. Svetitsky and L.D. McLerran,
Phys.Rev.{\bf D34},783(1986)
\bibitem{durand} L. Durand and E. Mendel, Phys.Rev.{\bf
D26},1368(1982)
\bibitem{rom}P. Romatschke, ArXiv:0902.3663













\end{thebibliography}
\end{document}